\documentclass[preprint,prd,nofootinbib,tightenlines,amsmath]{revtex4}
\usepackage{enumerate}
\usepackage{multirow}

\usepackage{epsfig}
\usepackage{graphics,color}      
\usepackage{verbatim}
\usepackage{amsmath}    
\catcode`\@=11
\def\sla@#1#2#3#4#5{{%
 \setbox\z@\hbox{$\m@th#4#5$}%
 \setbox\tw@\hbox{$\m@th#4#1$}%
 \dimen4\wd\ifdim\wd\z@<\wd\tw@\tw@\else\z@\fi
 \dimen@\ht\tw@
 \advance\dimen@-\dp\tw@ \advance\dimen@-\ht\z@
 \advance\dimen@\dp\z@
 \divide\dimen@\tw@ \advance\dimen@-#3\ht\tw@
 \advance\dimen@-#3\dp\tw@ \dimen@ii#2\wd\z@
 \raise-\dimen@\hbox to\dimen4{%
 \hss\kern\dimen@ii\box\tw@\kern-\dimen@ii\hss}%
 \llap{\hbox to\dimen4{\hss\box\z@\hss}}}}

\def\slashed#1{%
 \expandafter\ifx\csname sla@\string#1\endcsname\relax
{\mathpalette{\sla@/00}{#1}}
\fi}
\def\declareslashed#1#2#3#4#5{%
 \expandafter\def\csname sla@\string#5\endcsname{%
#1{\mathpalette{\sla@{#2}{#3}{#4}}{#5}}}}
 \catcode`\@=12
\declareslashed{}{/}{.08}{0}{D}
 \declareslashed{}{/}{.1}{0}{A}
 \declareslashed{}{/}{0}{-.05}{k}
 \declareslashed{}{/}{.1}{0}{\partial}
 \declareslashed{}{\not}{-.6}{0}{f}

\def\lsim{\mathrel {\vcenter {\baselineskip 0pt \kern 0pt
    \hbox{$<$} \kern 0pt \hbox{$\sim$} }}}
\def\gsim{\mathrel {\vcenter {\baselineskip 0pt \kern 0pt
    \hbox{$>$} \kern 0pt \hbox{$\sim$} }}}

\newcommand{\bea}{\begin{eqnarray}}
\newcommand{\eea}{\end{eqnarray}}

\begin{document}

\baselineskip=15pt
\preprint{}

\title{Constraints on anomalous color dipole operators from Higgs boson production at the LHC}

\author{Alper Hayreter and German Valencia}

\email{valencia@iastate.edu}

\affiliation{Department of Physics, Iowa State University, Ames, IA 50011.}

\date{\today}

\vskip 1cm
\begin{abstract}

Physics beyond the standard model (SM) can be parameterized with  an effective Lagrangian that respects the symmetries of the standard model and contains many operators of dimension six. We consider  the subset of these operators that is responsible for flavor diagonal anomalous color magnetic (CMDM) and electric (CEDM) dipole couplings between quarks and gluons. Invariance of these operators under the SM implies that they contribute  to Higgs boson production at the LHC and we study the corresponding constraints that can be placed on them. For the case of the top-quark we first review constraints from top-quark pair production and decay, and then compare them to what can be achieved by studying $t\bar{t}h$ production. We also constrain the corresponding couplings for $b$-quarks and light quarks by studying $pp \to b\bar{b}h$ and $pp \to hX$ respectively.

\end{abstract}

\pacs{PACS numbers: }

\maketitle

\section{Introduction}

The flavor diagonal dipole couplings between quarks and gluons of magnetic and electric type are generalizations of the electric dipole moment and anomalous magnetic moment of fermions given by the dimension five operators,
\begin{eqnarray}
{\cal L}=\frac{g_s}{2}\ \bar{f}\ T^a\sigma^{\mu\nu}\left(a_f^g+i\gamma_5 d_f^g\right) \ f\ G^a_{\mu\nu}.
\label{defcoup}
\end{eqnarray}
The operator in Eq.~\ref{defcoup} is not gauge invariant under the full standard model  gauge group and this needs to be remedied in the context of effective field theories which provide 
a general framework to study physics beyond the  SM. Within this formalism one assumes that there is some new physics (NP) that appears at a high energy scale $\Lambda$ that can be integrated out and represented by an effective Lagrangian valid for energies below $\Lambda$. The dominant effects of the new physics at scales below $\Lambda$ are then described by the lowest dimension operators, and complete catalogs for these operators (at least up to dimension six) exist in the literature \cite{Buchmuller:1985jz,Grzadkowski:2010es}.

After last year's discovery of a light Higgs boson with mass $m_h \sim 126$~GeV \footnote{Interpreting this state to be an elementary scalar.}, the gauge invariance of the effective Lagrangian must be imposed within the framework of the linear realization of electroweak symmetry breaking. The effective Lagrangian thus includes the Higgs field as an active degree of freedom. This has  implications for the operators in Eq.~\ref{defcoup}: they really are of dimension six and they contain couplings to Higgs bosons. The explicitly gauge invariant form is given by (in the notation of Ref.~\cite{Buchmuller:1985jz})
\begin{eqnarray}
{\cal L} = g_s\frac{d_{uG}}{\Lambda^2}\ \bar{q}\sigma^{\mu\nu}T^au\  \tilde\phi G^a_{\mu\nu} + g_s\frac{d_{dG}}{\Lambda^2}\ \bar{q}\sigma^{\mu\nu}T^a d  \ \phi G^a_{\mu\nu} \  +\ {\rm h.c.}
\label{GIanocoup}
\end{eqnarray}
where $q$ is the SM quark doublet, $\phi$ is the scalar doublet, $\tilde\phi_i =\epsilon_{ij}\phi_j$ and the $SU(3)$ generators are normalized as ${\rm Tr}(T_aT_b)=\delta_{a,b}/2$. Electroweak symmetry is spontaneously broken when the scalar acquires a vacuum expectation value (vev) $<\phi>=v/\sqrt{2}$, $v\approx 246$~GeV. An expansion of this gauge invariant Lagrangian reproduces the terms in Eq.\ref{defcoup} with the identification
\begin{eqnarray}
a_{u,d}^g &=& \frac{\sqrt{2} \ v}{\Lambda^2}{\rm ~Re}(d_{uG,dG})\nonumber \\
d_{u,d}^g &=& \frac{\sqrt{2} \ v}{\Lambda^2}{\rm ~Im}(d_{uG,dG}).
\label{anomcouplings}
\end{eqnarray}
The effect of imposing invariance under the SM group is then that the  operators of Eq.~\ref{defcoup} can only exist in conjunction with  corresponding operators that involve a Higgs field. This opens up the possibility of constraining the couplings by studying processes with a Higgs boson, such as $pp \to f \bar{f}h$ which we do in this paper. In all cases, when we quote constraints on $d_{qG}$ we will implicitly assume a new physics scale $\Lambda=3$~TeV. 

\section{Case of the top-quark}

The flavor diagonal dipole couplings of the top-quark to gluons are known in the literature as the CEDM and the CMDM   and have been studied at length in connection to top-quark pair production and decay. The notation, however is not uniform and different definitions exist~\cite{Atwood:1992vj,Cheung:1995nt,Choi:1997ie,Sjolin:2003ah,Antipin:2008zx, Hioki:2009hm,HIOKI:2011xx,Hioki:2012vn,Gupta:2009wu,Biswal:2012dr}. A typical result is that of Ref.~\cite{Gupta:2009wu} where it is found that using CP-odd observables the 5$\sigma$ statistical sensitivity with 10~fb$^{-1}$ to $d^g_t$ is of order $0.1/m_t$. The CMDM has also been studied before \cite{Atwood:1992vj,Cheung:1995nt,Martinez:2007qf,Larkoski:2010am,Englert:2012by} and constraints have been derived from production cross-sections. Very recently Ref.~\cite{Baumgart:2012ay} has found that the study of spin correlations over the life span of the LHC can produce limits at the level of $0.03/m_t$ for both $a_t^g$ and $d_t^g$. 

We begin this section revisiting the constraints from 
top-quark pair production cross-sections and T-odd asymmetries both at 8 and 14~TeV. These will serve as comparison points for our study of constraints from $t\bar{t}h$ processes, a possibility that has also been known for a long time \cite{DeRujula:1990db} and has been revisited recently \cite{Choudhury:2012np,Degrande:2012gr}. Our results for the top-quark couplings are in agreement with these recent papers and we expand on them by considering the T odd correlations that would allow one to separate the effect of a CEDM from that of a CMDM and by extending the study to include light quarks.

\subsection{$t\bar{t}$ production and decay}

We first consider $t\bar{t}$ production  at the LHC at 8 TeV in order to compare with recent experimental results. Both CMS  \cite{CMS:2012fza} and ATLAS  have results that have not yet been combined. Since all the existing results are compatible, we will use one of them to illustrate the constraints on the new couplings. Taking the ATLAS cross-section in the lepton plus jets channel \cite{atlastop} with errors added in quadrature (but dominated by systematic error)
\begin{eqnarray}
\sigma(t\bar{t}) &=& (241\pm 32){\rm ~pb},
\end{eqnarray}
and comparing with the theoretical expectation obtained from HATHOR  \cite{Aliev:2010zk} as quoted by ATLAS,
\begin{eqnarray}
\sigma(t\bar{t}) &=& (238^{+22}_{-24}){\rm ~pb}
\end{eqnarray}
we find a representative ratio between the measured and expected cross-sections
\begin{eqnarray}
\frac{\sigma(t\bar{t})_{Exp}}{\sigma(t\bar{t})_{TH}} = 1.01 \pm 0.17.
\label{exp8}
\end{eqnarray}
In this section we will interpret the error in this ratio as the room that remains for new physics that can affect top-quark pair production.

We begin our calculation by implementing the Lagrangian of Eq.~\ref{GIanocoup} into {\tt MadGraph5} \cite{MadGraph} with the aid of {\tt FeynRules} \cite{Christensen:2008py}. We use the resulting model UFO file to generate top-quark pair events for different values of $a_t^g$ and $d_t^g$ in a range that changes the SM cross-section by factors of a few. We then fit  the numerical results to obtain approximate expressions for cross-sections and asymmetries in terms of  the new couplings. The results of our simulations are presented in the Appendix in Figure~\ref{dtgsig}, and the corresponding fit to these points is given in Eq.~\ref{fit8}. 

\begin{figure}[thb]
\includegraphics[width=0.6\textwidth,]{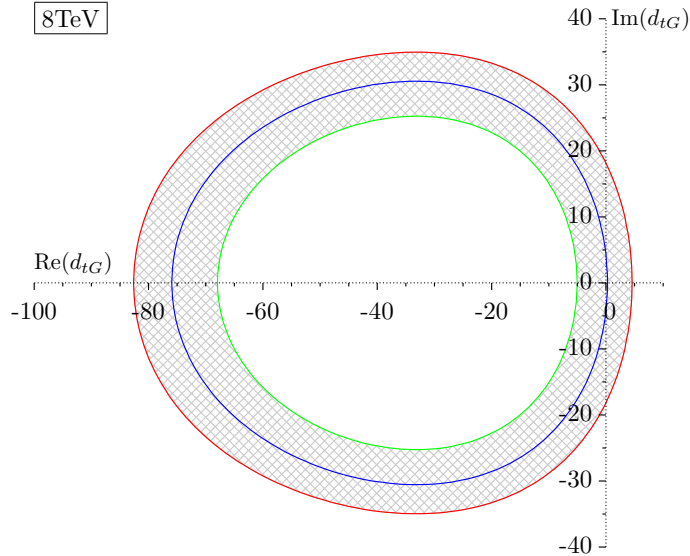}
\caption{$d_{tG}$ parameter space allowed by the measurement of the $t\bar{t}$ production cross-section at the LHC at 8 TeV.  The center curve (blue) corresponds to the central value of Eq.~\ref{exp8}, and the shaded region is that allowed at 1$\sigma$.}
\label{f:contour8}
\end{figure}

In Figure~\ref{f:contour8} we compare the fit to the current measurement in the form of Eq.~\ref{exp8},
and from this comparison we extract the current LHC exclusion region. The central value (for which the measured cross-section is approximately equal to the SM) is reproduced along the curve that goes through the origin. Along this curve there is a cancellation between pure NP contributions and interference between NP and the SM. The shaded region shows the parameters allowed by Eq.~\ref{exp8} at the $1\sigma$ level. 

Constraining these couplings one at a time, results in the $1\sigma$ ranges for the top CMDM and CEDM 
\begin{eqnarray}
-0.034 \lsim m_t a_t^g  \lsim 0.031&{\rm or}&-0.55 \lsim m_t a_t^g  \lsim -0.46 ,\nonumber \\
&{\rm and}&  |d_t^g |\lsim  0.12/m_t.
\label{8tevresults}
\end{eqnarray}
From Figure~\ref{f:contour8} we see that there are two allowed ranges for the real part of $d_{tG}$ (proportional to $a_t^g$). Of course the effective Lagrangian formalism assumes that the new physics is small compared to the SM and this makes the range closest to the origin, presented in Table~\ref{t:results}, the more natural one. Approximately, this is the range one would get by keeping only the term linear in the CMDM in Eq.~\ref{exp8}. This figure is in agreement with that of Ref.~\cite{Hioki:2012vn} with an opposite sign convention for $d_{tG}$.

Next we consider possible improvements at 14~TeV. It should be clear that simply measuring the $t\bar{t}$ production cross-section will not change the above picture much as long as the uncertainty in the cross-section is dominated by systematics. More interesting is the possibility of using large statistics samples at 14~TeV to  measure  asymmetries in specific channels. In particular T-odd correlations are linear in the CEDM and potentially have a better sensitivity to this type of new physics.  Here we will focus on the clean di-muon channel but generalizations to other channels are possible and have been discussed before \cite{Gupta:2009eq}. For  our numerical study we use default {\tt MadGraph5} cuts that require the intermediate top-quarks and $W$ bosons in the process $pp\to t\bar{t}\to b\bar{b}\mu^+\mu^- \nu\bar\nu$ to be within 15 widths of their mass shell. We use MSTW2008 parton distribution functions and a set of acceptance and separation cuts for muons and $b$ quarks given by
\begin{eqnarray}
p_T(\mu^\pm) > 20~{\rm GeV}&& p_T(b,{\bar b}) > 25 ~{\rm GeV} 
\nonumber \\
|\eta(b,{\bar b},\mu^\pm)| < 2.5 && \Delta R (b{\bar b}) > 0.4.
\label{cuts1}
\end{eqnarray}
In addition we use a missing energy requirement, $\slashed{E}_T > 30$~GeV, that is known to significantly reduce the background. 

\begin{figure}[thb]
\includegraphics[width=0.6\textwidth,]{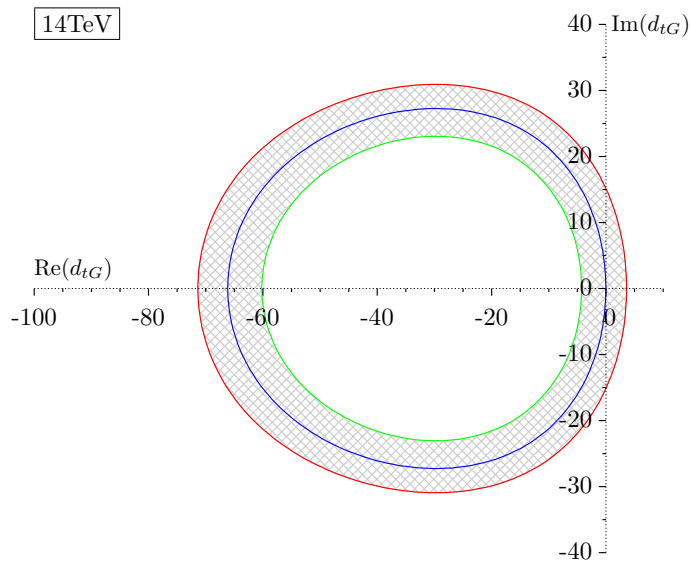}
\caption{$d_{tG}$ parameter space for which the predicted $t\bar{t}$ production cross-section at the LHC at 14 TeV is within the $1\sigma$ theoretical error in the SM cross-section at NLO.}
\label{f:contour14}
\end{figure}

As detailed in the Appendix, we estimate the dependence of the cross-section on $d_{tG}$ by generating event samples at different values of the CEDM and CMDM and extrapolating with a fit, Eq.~\ref{fit14}. This fit is shown in Figure~\ref{f:contour14}, where the central (blue) curve corresponds to a cross-section matching the central value of the SM prediction. The shaded region shows the parameter space for which the NP contributions are within the  1~$\sigma$ errors in the {\it theoretical}~ SM cross-section at next-to-leading-order (NLO), $\sigma=884^{+125}_{-121}$~pb \cite{Beneke:2011ys} 
Once again we can summarize the results taking only one non-zero coupling at a time as 
\begin{eqnarray}
-0.029 \lsim m_t a_t^g  \lsim 0.024&{\rm or}&-0.480 \lsim m_t a_t^g  \lsim -0.478 ,\nonumber \\
&{\rm and}&  |d_t^g |\lsim  0.10/m_t.
\label{14tevresults}
\end{eqnarray}
As expected, this measurement will not result in a significant improvement over the bound with 8~TeV data, Eq.~\ref{8tevresults}, unless the experimental systematic errors and the theoretical errors in the SM can both be reduced to the few percent level. As before, we only include the more natural range for $a_t^g$ in Table~\ref{t:results}.

Next we consider a T-odd correlation that can single out the CP violating CEDM (a correlation that has been studied before). As opposed to CP even observables such as cross-sections, this asymmetry is linear in the CEDM and therefore more sensitive to it. To quantify our bounds  we consider the integrated asymmetry in the lab frame defined by
\begin{eqnarray}
{\cal A}&=&\frac{\sigma({\cal O}_1>0)-\sigma({\cal O}_1<0)}{\sigma_{SM}},
\label{Tasym}
\end{eqnarray}
associated with the correlation
\begin{eqnarray}
{\cal O}_1 &=& \vec{p}_t\cdot (\vec{p}_{\mu^+}\times\vec{p}_{\mu^-}) .
\label{corr1}
\end{eqnarray}
We choose Eq.~\ref{corr1}, as this is the T-odd correlation most sensitive to the CEDM out of the many  possible for this process \cite{Gupta:2009wu}. One correlation will suffice for our purpose of comparing the bounds that can be obtained from processes with and without Higgs bosons. Of course, a more realistic discussion of this correlation requires using momenta that can be reconstructed; discussion of backgrounds; and possible combinations of different channels. All these aspects have been discussed previously in the literature and we do not repeat that discussion.

Notice that we have defined the integrated asymmetry in Eq.~\ref{Tasym} using the SM cross-section in the denominator. This choice ensures that the asymmetry is linear in the CEDM and does not depend on the new physics indirectly, through the cross-section. Experimentally it is simpler to define a counting asymmetry which would correspond to Eq.~\ref{Tasym} with the cross-section as a function of the CEDM in the denominator. Such counting asymmetry is only approximately equal to ${\cal A}$ for small values of the CEDM. 
\begin{figure}[thb]
\includegraphics[width=0.45\textwidth]{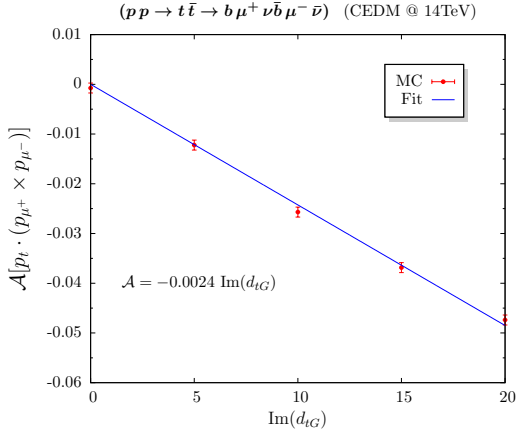}
\caption{Asymmetries in the process $pp\to t\bar{t}$ in the di-muon channel as calculated with {\tt MadGraph5} for different values of the anomalous coupling $d_{tG}$ with the cuts of Eq.~\ref{cuts1} and the corresponding fit.}
\label{dtgasym}
\end{figure}
We present our numerical results for the asymmetry for four different values of $ {\rm ~Im}(d_{tG})$, as well as a linear fit to these results in Figure~\ref{dtgasym}.  The fit indicates that a measurement with 10~fb$^{-1}$ would have a (1$\sigma$) statistical sensitivity of
\begin{eqnarray}
|d_t^g|& \lsim& 0.009/m_t,
\end{eqnarray}
an order of magnitude better than what can be achieved from a measurement of the cross-section, Eq.~\ref{14tevresults}. This sensitivity is comparable to what can be achieved for $a_t^g$ from cross-section measurements, reflecting the fact that the CMDM contributes linearly to the cross-section. In principle it is also possible to obtain increased sensitivity to $a_t^g$ using other observables such as spin correlations, but existing results in the literature suggest the improvement is only by factors of a few \cite{Baumgart:2012ay}.  

The best existing indirect constraint on the top CEDM arises from contributions to the neutron electric dipole moment (edm) \cite{DeRujula:1990db} (rescaled to the current limit \cite{Baker:2006ts}) via the Weinberg operator \cite{Weinberg:1989dx} giving $ |m_t d_t^g |\lsim  2.4 \times 10^{-4}$. Although the LHC will probably not reach this limit, direct searches are always complementary to indirect constraints which rely on multiple assumptions.

\subsection{$t\bar{t}h$ production}

We turn our attention to the Higgs production process in association with a top-quark pair.  We begin as usual with a study of the cross-section which we calculate numerically at leading-order (LO) with {\tt MadGraph5}. Our numerical results for several values of the top CEDM and the top CMDM couplings as well as a fit to these Monte Carlo (MC) points are presented in the Appendix in Figure~\ref{dtgsigh} and Eq.~\ref{sigtth}. We normalize our results to the SM cross-section and use them to estimate the sensitivity that can be obtained at $\sqrt{S}=14$~TeV. 

\begin{figure}[thb]
\includegraphics[width=0.6\textwidth,]{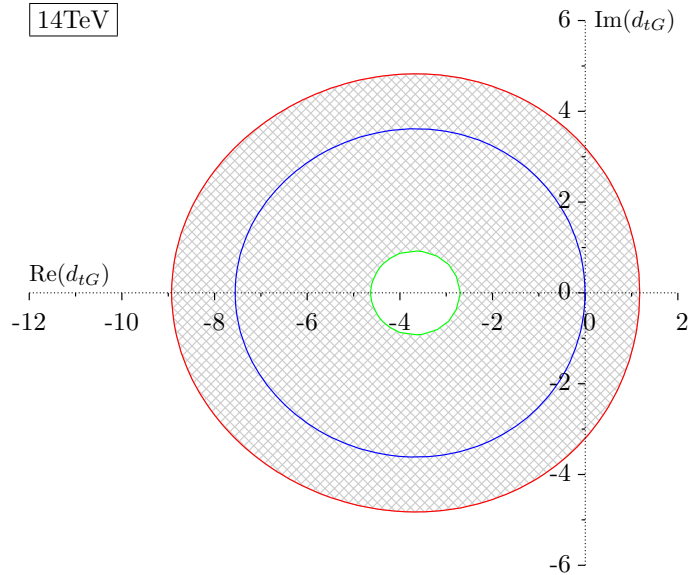}
\caption{$d_{tG}$ parameter space as it would be constrained by a measurement of the $t\bar{t}h$ production cross-section at the LHC at 14 TeV.  The central (blue) curve corresponds to cross-sections that are equal to the central value of the SM. The shaded region is the one in which the NP contributions are below the  $1\sigma$ theoretical error for the SM at NLO.}
\label{f:contour_tth14}
\end{figure}

In Figure~\ref{f:contour_tth14}  we illustrate the region in the $a_t^g-d_t^g$ parameter space that would be allowed by assuming that the central value of a future measurement agrees with the SM value (blue curve). The shaded region corresponds to that in which the new physics effects modify the SM cross-section by at most 1$\sigma$ using the theoretical errors in the NLO cross-section for the SM, $\sigma(pp\to t\bar{t}h)_{NLO}= (611^{+92}_{-110})$~fb \cite{Beenakker:2001rj,Dawson:2002tg,Dittmaier:2011ti}. For $t\bar{t}h$ we only consider the 14~TeV LHC because the existing 8~TeV results for this mode are only sensitive to cross-sections about 5-6 times larger than the SM  \cite{Chatrchyan:2013yea}. 
Assuming that the experimental error can be kept below the theoretical uncertainty at NLO a future measurement can reach a 1-$\sigma$ sensitivity to the ranges
\begin{eqnarray}
-0.06 \lsim m_t a_t^g  \lsim 0.03&{\rm or}&-0.016 \lsim m_t a_t^g  \lsim 0.008 ,\nonumber \\
&{\rm and}&  |d_t^g |\lsim  0.02/m_t.
\label{14tevresultstth}
\end{eqnarray}
As before, we have only included the more natural range for $a_t^g$ of the two allowed solutions in Table~\ref{t:results}. We see a significant improvement over what can be achieved with the top-quark pair process, Eq.~\ref{14tevresults} for the CEDM and a modest improvement for the CMDM. 

\begin{figure}[thb]
\includegraphics[width=0.45\textwidth,]{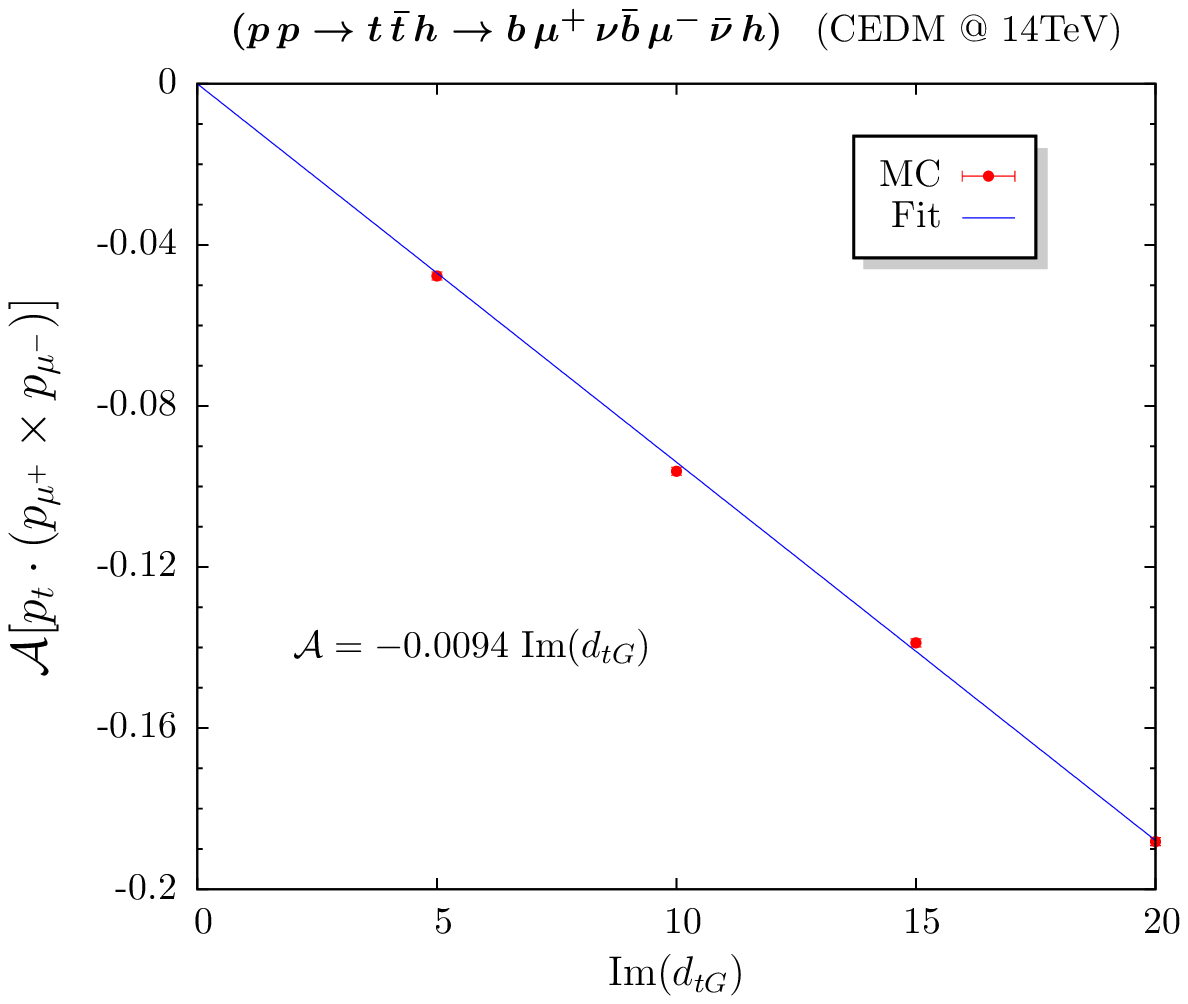} \hspace{1cm}
\includegraphics[width=0.45\textwidth,]{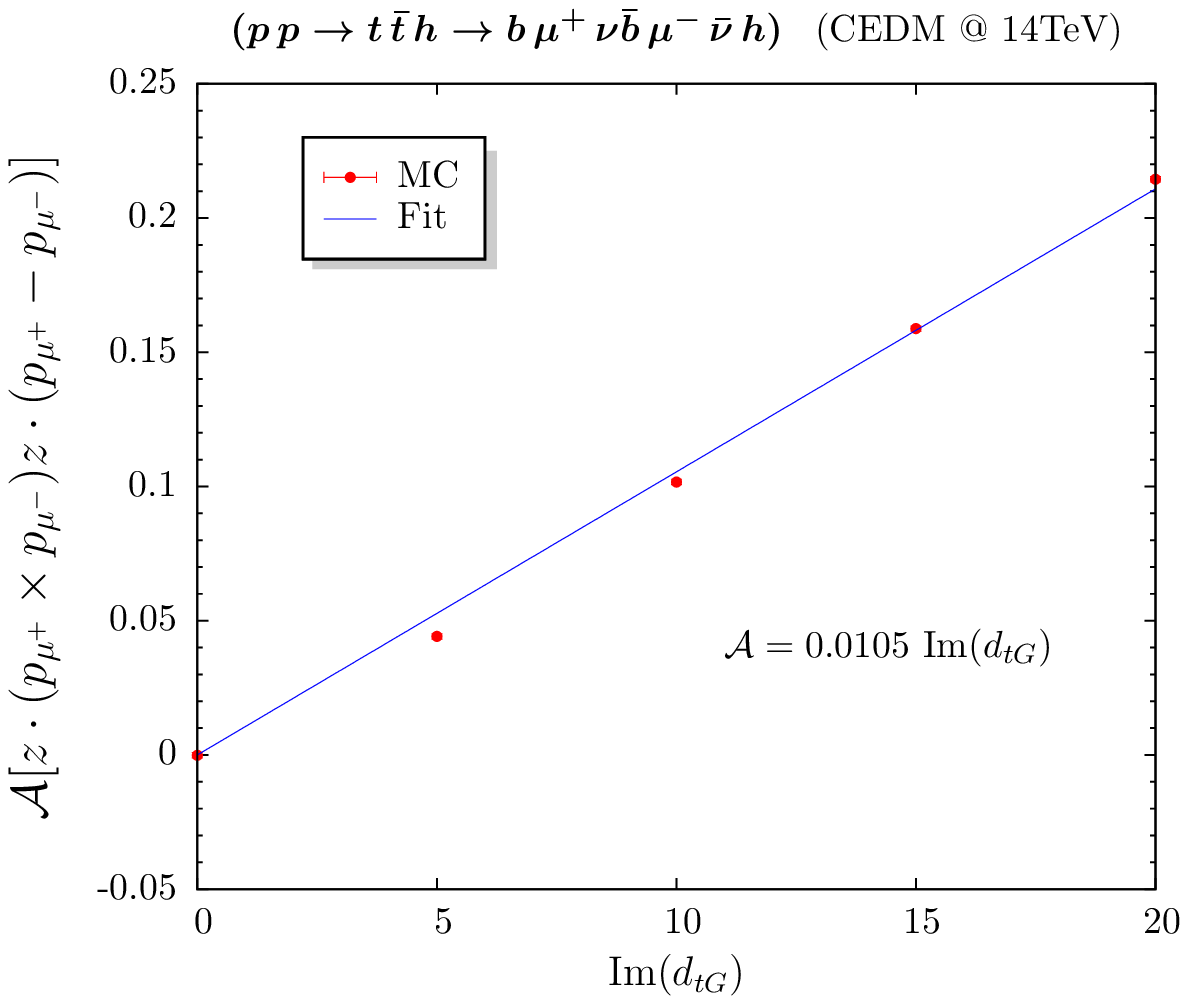}
\caption{T-odd asymmetries with the largest sensitivity to the top quark CEDM in the process $t\bar{t}h$. The points correspond to {\tt MadGraph5} simulations for four different values of the CEDM, and the line is a linear fit to these points.}
\label{f:tth-asym}
\end{figure}

Next we can try to isolate the CP violating coupling ${\rm ~Im}(d_{tG})$ with the use of the T-odd asymmetries $A_{1,2}$, associated with the correlation ${\cal O}_{1}$ of Eq.~\ref{corr1} but for the process $pp\to t\bar{t}h \to b\bar{b}h\mu^+\mu^-\nu\bar\nu$, and with a second correlation
\begin{eqnarray}
{\cal O}_2 &=& \vec{p}_{beam}\cdot (\vec{p}_{\mu^+}\times\vec{p}_{\mu^-})\  \vec{p}_{beam}\cdot (\vec{p}_{\mu^+}-\vec{p}_{\mu^-}).
\label{corr2}
\end{eqnarray}
In principle this process admits additional correlations involving the Higgs momenta. Numerically, however, we find all of them to be significantly smaller. For our study of the asymmetries we assume that the top quarks decay as in the SM and consider only the di-muon channel in order to construct them, keeping in mind that the statistics can be increased by using other channels. No attempt is made to include the Higgs decay or the complications involved with its reconstruction. The results of our MC simulation for a few values of the top CEDM as well as a linear fit to those points are shown in Figure~\ref{f:tth-asym}.
Using the central value of the NLO cross-section for $pp\to t\bar{t}h$ at 14~TeV in the SM,  611~fb \cite{Dittmaier:2011ti}, we see that to measure percent level asymmetries in this channel at the 1$\sigma$ level would require about a thousand fb$^{-1}$. When these integrated luminosities have been collected the asymmetries in Figure~\ref{f:tth-asym} could reach a sensitivity to $m_t d_t^g$ at the $0.007$ level. Obviously one can improve the statistics significantly by considering other top decay channels, but the number of $t\bar{t}h$ events  is small so this type of measurement will not be possible any time soon.

\section{Case of the bottom quark}

We now turn our attention to the case of the $b$-quark anomalous CEDM and CMDM couplings. In this case the $b\bar{b}h$ process is possibly the only handle on these couplings at the LHC as 
the $bb$ cross-section is completely dominated by QCD and the contributions of the new couplings are negligible. The production of $b\bar{b}$ pairs in association with a Higgs boson has been studied  before, with a NLO result for $m_H=120$~GeV of \cite{Dittmaier:2003ej}
\begin{eqnarray}
\sigma(pp\to b\bar{b}hX)_{SM}&=&(5.8\pm1.0)\times 10^2{\rm ~fb}.
\end{eqnarray}
More recently this mode has been studied in connection to Higgs bosons in the MSSM where the $b\bar{b}h$ coupling is enhanced for large $\tan\beta$ \cite{mssmbbh}.

As before, we generate MC samples for several values of the CEDM and CMDM of the $b$-quark chosen so that they change the SM cross-section by factors of a few, and we then perform a fit to these values. We generate the $b\bar{b}h$ events with default {\tt MadGraph5} cuts, without $h$ decay, and without backgrounds. These results are presented  in the Appendix in Figure~\ref{dbgsigh} and Eq.~\ref{sigbbh} respectively. 

\begin{figure}[thb]
\includegraphics[width=0.6\textwidth,]{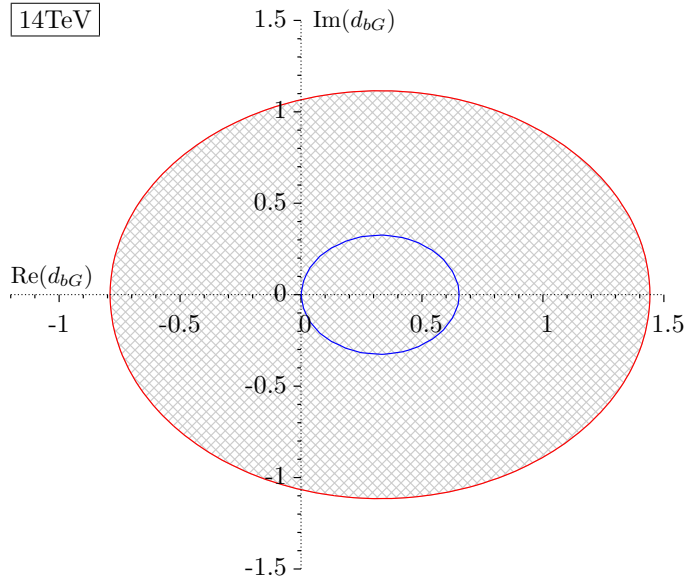}
\caption{$d_{bG}$ parameter space allowed by a measurement of  the $b\bar{b}h$ production cross-section at the LHC at 14 TeV. The shaded area shows the region in which the NP contributions are below the $1\sigma$ theoretical error in the NLO SM result.}
\label{f:contour_bbh14}
\end{figure}

Our fit can then be used to estimate the statistical sensitivity to the b-quark CEDM and CMDM that can be achieved with a measurement of the $b\bar{b}h$ cross-section. In Figure~\ref{f:contour_bbh14} we illustrate the region that would be allowed by requiring the new physics to remain within 1$\sigma$ of the NLO theoretical cross-section. The central curve (blue) corresponds to values of $d_{bG}$ which reproduce the central value of the SM cross-section. 
We can extract from the figure the $1\sigma$ ranges with only one coupling at a time (using $m_b= 4.2$~GeV)
\begin{eqnarray}
-1.3 \times 10^{-4} \lsim m_b a_b^g  \lsim 2.4 \times 10^{-4} &  &   |d_b^g |\lsim  (1.7 \times 10^{-4})/m_b.
\label{14tevresultsbbh}
\end{eqnarray}
Unlike the case of the top-quark, the interference between the b-quark CMDM and the SM is very small so that its effect on the cross-section is dominated by a quadratic term, Eq.~\ref{sigbbh}. For this reason there is only one range for the allowed $a_b^g$ in Eq.~\ref{14tevresultsbbh}. 

The best existing indirect constraint on the b-quark CEDM arises from contributions to the neutron edm \cite{DeRujula:1990db} via the Weinberg operator \cite{Weinberg:1989dx}. Rescaling to the current limit on the neutron edm, we find $|m_b d_b^g |\lsim  2 \times 10^{-8}$.

\section{The light quarks including charm}

For LHC processes involving only the light quarks, the effects of anomalous $a_q^g$ or $d_q^g$ couplings are buried in QCD background and Higgs production offers a unique window into the color dipole operators at high energy. 

The light quark CEDM and CMDM couplings contribute to  Higgs production at the LHC dominantly through the parton level process $q g \to q h$ and to a lesser extent $q \bar{q} \to h g$. These processes are dominated by the charm-quark within the SM. There are several implications from this observation:

\begin{itemize}

\item We can look for the effect of the new couplings in the LO Higgs production process $pp\to hX$. Our strategy to constrain the new couplings in this case, is to require the new contributions to fall within the theoretical uncertainty of the dominant SM gluon fusion cross-section and we pursue this route below. 
\item The SM tree-level contribution to Higgs production from $q g \to q h$ and  $q \bar{q} \to h g$ for the light quarks is very small and dominated by the charm quark. This implies that any interference between the SM and the CMDM is negligible except for charm. The new physics effects are thus purely quadratic in both the CEDM and the CMDM of the light quarks in our present study. This picture fails beyond LO, where $q g \to q h$ proceeds through a heavy quark loop \cite{Field:2003yy}, resulting in a more sizable contribution from light quarks which can interfere with the new physics. 
\item The new physics contribution could also be studied in the context of Higgs plus one jet (or more) production. This path would have several advantages: 
\begin{enumerate}
\item the NP would be compared to a smaller SM cross-section 
\item the interference between the SM and the CMDM would be included, resulting in a linear dependence on NP. 
\item the NP/SM ratio increases with $p_{T}$ so requiring a jet is favorable for the purpose of constraining NP.
\end{enumerate}
At the same time, Higgs plus jets production is a much more complicated process \cite{Field:2003yy,Campbell:2012am} and its study is beyond the scope of this paper. We will content ourselves for now with the simpler $pp\to hX$ study, keeping in mind that NLO processes may become much more relevant if there are hints for NP.
\end{itemize}

In view of these considerations we turn again to {\tt MadGraph5} to estimate the cross-section $pp\to hX$  due to new physics for different values of the anomalous couplings. We begin with $d_{uG}$
at 8~TeV for which we find 
\begin{eqnarray}
\sigma(pp\to hX) &\approx & 2.3 +0.176\left[{\rm Re}(d_{uG})^2+{\rm Im}(d_{uG})^2\right]{\rm ~pb}.
\label{pphxfit}
\end{eqnarray}
The SM contribution here, 2.3~pb, arises primarily from $c\bar{c}\to Hg$ and doesn't interfere with the NP which affects only the up-quark. As argued above, we ignore this contribution and compare the NP directly with 
the dominant gluon-fusion $pp\to hX$ cross-section for $m_H=125.5$~GeV \cite{gluonfusion}
\begin{eqnarray}
\sigma(pp\to hX) = (19.37\pm14.7){\rm ~pb}.
\end{eqnarray}
Requiring that the new physics be smaller than the theoretical error leads to the constraint
\begin{eqnarray}
|a_u^g|,|d_u^g| \lsim 3.5 \times 10^{-4} {\rm ~GeV}^{-1}.
\end{eqnarray}

We repeat the exercise for 14~TeV, this time including all the light quarks.  We find
\begin{eqnarray}
\sigma(pp\to hX) &\approx & 5.58 +0.7\left[{\rm Re}(d_{uG})^2+{\rm Im}(d_{uG})^2\right] \nonumber \\
&+& 0.4 \left[{\rm Re}(d_{dG})^2+{\rm Im}(d_{dG})^2\right]+0.1 \left[{\rm Re}(d_{sG})^2+{\rm Im}(d_{sG})^2\right] \nonumber \\
&-& 0.034 {\rm ~Re}(d_{cG})+0.07 \left[{\rm Re}(d_{cG})^2+{\rm Im}(d_{cG})^2\right]
\label{fithj14}
\end{eqnarray}
This is then compared to the NLO theoretical (dFG) cross-section and error as given in Ref.~\cite{Dittmaier:2011ti} for a mass of $M_H=125$~GeV
\begin{eqnarray}
\sigma(pp\to hX) = (49.97^{+7.3}_{-7.0}){\rm ~pb}
\label{nlo14}
\end{eqnarray}
by requiring the new physics contributions to fall within the $1\sigma$ uncertainty.
\begin{figure}[thb]
\includegraphics[width=0.6\textwidth,]{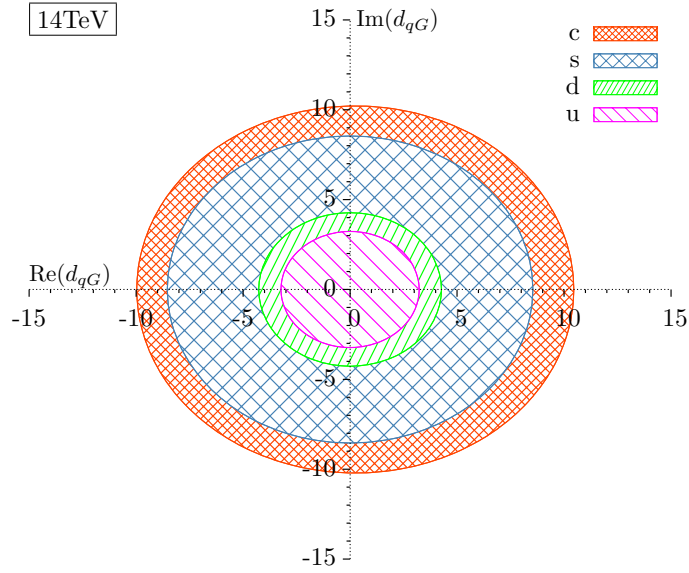}
\caption{$pp\to HX$ production cross-section at the LHC at 14 TeV  as a function of the light quarks CEDM and CMDM couplings allowing only one $d_{qG}$ to be nonzero at a time. The  concentric regions, starting with the smallest one, show the  allowed couplings at $1\sigma$ for $u,d,s$ and $c$ quarks respectively.}
\label{f:contour_hj}
\end{figure}

The resulting allowed regions, within the 1$\sigma$ range of Eq.~\ref{nlo14}, and allowing only one $d_{qG}$ to be nonzero at a time are shown in Figure~\ref{f:contour_hj}.  For the up, down and strange quarks the allowed regions are just circles of larger radii starting with the inner most for up ending with the one for strange. The outermost region is for charm and shows a slight asymmetry due to the interference with the SM. In terms of CEDM and CMDM couplings these limits correspond to
\begin{eqnarray}
|{a}_u^g|,|{d}_u^g| \lsim  1.2\times 10^{-4}{\rm ~GeV}^{-1} &&|{a}_d^g|,|{d}_d^g| \lsim  1.6\times 10^{-4}{\rm ~GeV}^{-1}\nonumber \\
|{a}_s^g|,|{d}_s^g| \lsim  3.3 \times 10^{-4}{\rm ~GeV}^{-1}   && |{d}_c^g| \lsim 3.9 \times 10^{-4}{\rm ~GeV}^{-1}\nonumber \\
-3.8 \times 10^{-4}{\rm ~GeV}^{-1} \lsim & |a_c^g| &\lsim   4.0 \times 10^{-4}{\rm ~GeV}^{-1}
\end{eqnarray}

We are only aware of existing low energy constraints for the CEDM,  
where the strongest bound arises from the neutron electric dipole moment ($d^\gamma_N$).
We can estimate the contribution of an up-quark (down-quark) CEDM to the neutron edm using  naive dimensional analysis \cite{DeRujula:1990db,Pospelov:2005pr} as
\begin{eqnarray}
d^\gamma_N &=&\frac{e}{4\pi} d_{u,d}^g
\end{eqnarray}
and from the experimental limit on the neutron edm, $d^\gamma_N < 2.9 \times 10^{-26}$~e-cm \cite{Baker:2006ts} we then find
\begin{eqnarray}
d_{u,d}^g&\lsim& 1.8\times 10^{-11}{\rm ~GeV}^{-1}.
\end{eqnarray}
Similarly from the limit on the $\Lambda$ edm, $d^\gamma_\Lambda < 1.5 \times 10^{-16}$~e-cm \cite{Pondrom:1981gu} we find 
\begin{eqnarray}
d_{s}^g&\lsim& 0.1 {\rm ~GeV}^{-1}.
\end{eqnarray}
Using the contribution from the Weinberg operator to the neutron edm, and with $m_c \sim 1.28$~GeV, we also obtain
\begin{eqnarray}
m_c d_{c}^g&\lsim& 6 \times 10^{-10}.
\end{eqnarray}

\section{Conclusions}

We have studied the constraints that can be obtained on CEDM and CMDM type couplings of quarks by studying Higgs boson production at the LHC. For the case of the top quark we have compared them to what can be obtained from studying top-quark pair production and decay. We find that the $pp\to t\bar{t}h$ cross-section can provide a significant improvement on the CEDM bound over that from the $pp\to t\bar{t}$ cross-section, and comparable to what can be obtained using T-odd asymmetries in the latter. A further improvement is in principle possible by studying T-odd correlations in $pp\to t\bar{t}h$. The low statistics for this channel imply that thousands of fb$^{-1}$ would be needed. The improvement on the top CMDM constraint is modest, on the other hand. 

For the other quarks, the Higgs production process may be the only opportunity to constrain them at LHC. Our results are summarized in Table~\ref{t:results}.
\begin{table}[h]
\caption{Summary of results for 1$\sigma$ bounds that can be placed on the CEDM and CMDM couplings of quarks at the LHC. The last column shows the effective new physics scale than can be probed by the LHC with the given process, the two numbers corresponding to the CMDM and the CEDM respectively.}
\begin{center}
\begin{tabular}{|c|c|c|c|}
\hline
Process & CMDM & CEDM & $\Lambda$ (TeV) \\ \hline
$\sigma(pp \to t\bar{t})$ 8 ~TeV& $-0.034\lsim m_ta_t^g\lsim 0.031$ &  $|m_t d_t^g|\lsim 0.12$ & (1.5, .7)\\  \hline
$\sigma(pp \to t\bar{t})$ 14 ~TeV& $-0.029\lsim m_ta_t^g\lsim 0.024$ &  $|m_t d_t^g|\lsim 0.1$ & (1.5, .7)\\  \hline
$A_1(pp \to t\bar{t})$ 14 TeV&- &  $|m_t d_t^g|\lsim 0.009$ &(-, 2.5) \\  \hline
$\sigma(pp \to t\bar{t}h)$ 14~TeV& $-0.016\lsim m_ta_t^g\lsim 0.008$ &  $|m_t d_t^g|\lsim 0.02$ & (2, 1.7) \\  \hline
$A_{1,2}(pp \to t\bar{t}h)$ 14~TeV& - &  $|m_t d_t^g|\lsim 0.007$ &(-, 3) \\  \hline
$\sigma(pp \to b\bar{b}h)$ 14~TeV& $-1.3 \times 10^{-4}\lsim m_ba_b^g\lsim 2.4\times 10^{-4}$ &  $|m_b d_b^g|\lsim 1.7 \times 10^{-4}$& 2.7  \\  \hline
$\sigma(pp \to hX)$ 8~TeV& $|a_u^g |\lsim 3.5 \times 10^{-4}\ {\rm GeV}^{-1}$ &  $|d_u^g |\lsim 3.5 \times 10^{-4}\ {\rm GeV}^{-1}$ & 1 \\  \hline
$\sigma(pp \to hX)$ 14~TeV& $|a_u^g |\lsim 1.2 \times 10^{-4}\ {\rm GeV}^{-1}$ &  $|d_u^g |\lsim 1.2 \times 10^{-4}\ {\rm GeV}^{-1}$ &1.7 \\  \hline
$\sigma(pp \to hX)$ 14~TeV& $|a_d^g |\lsim 1.6 \times 10^{-4}\ {\rm GeV}^{-1}$ &  $|d_d^g |\lsim 1.6 \times 10^{-4}\ {\rm GeV}^{-1}$ &1.5 \\  \hline
$\sigma(pp \to hX)$ 14~TeV& $|a_s^g |\lsim 3.3 \times 10^{-4}\ {\rm GeV}^{-1}$ &  $|d_s^g |\lsim 3.3 \times 10^{-4}\ {\rm GeV}^{-1}$ &1 \\  \hline
$\sigma(pp \to hX)$ 14~TeV& $|a_c^g |\lsim 3.9 \times 10^{-4}\ {\rm GeV}^{-1}$ &  $|d_c^g |\lsim 3.9 \times 10^{-4}\ {\rm GeV}^{-1}$ &1 \\  \hline
\end{tabular}
\end{center}
\label{t:results}
\end{table}
In this Table we show the constraints we obtained on the anomalous color electric and magnetic dipole moment couplings of quarks in processes with Higgs bosons (and without for the case of top). In the last column we use Eq.~\ref{anomcouplings} to interpret these limits as the scale of new physics that can be probed by these processes at LHC under the usual assumption that the effective Lagrangian dimensionless couplings $d_{qG}$ are of order one. 

\begin{acknowledgments}

This work was supported in part by the DOE under contract number
DE-FG02-01ER41155. 

\end{acknowledgments}

\appendix

\section{$pp\to t\bar{t}$ events}

\begin{figure}[thb]
\includegraphics[width=0.45\textwidth]{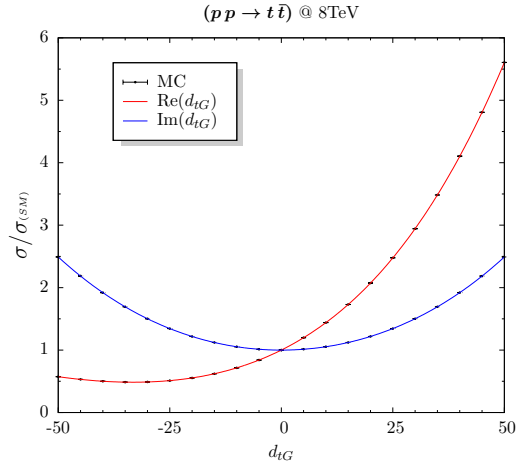}
\caption{Cross-section for the process $pp\to t\bar{t}$ at $\sqrt{S}=8$~TeV as calculated with {\tt MadGraph5} for different values of the anomalous coupling $d_{tG}$ and the corresponding quartic fit described in the text.}
\label{dtgsig}
\end{figure}

We first use {\tt MadGraph5} to compute the $t\bar{t}$ production cross-section at the LHC at 8 TeV for a set of values of  ${\rm Im}(d_{tG})$ or ${\rm Re}(d_{tG})$ with $\Lambda=3$~TeV. The results of these numerical calculations, including the error estimated by {\tt MadGraph5}, are plotted in Figure~\ref{dtgsig}. In the same figure we show the result of a quartic fit to these points that is guided by theoretical prejudice:  we know that the cross-section will be a quartic polynomial in the anomalous couplings because they occur once ($q\bar{q}$ annihilation and s-channel gluon fusion diagrams) or twice (t-and-u-chanel gluon fusion diagrams) at the amplitude level. We also know that the CP violating coupling ${\rm Im}(d_{tG})$ can only contribute to the cross-section through even powers. Finally, from an analytic calculation of the squared matrix element for the gluon fusion process, we know that the quartic terms in ${\rm Re}(d_{tG})$ and in ${\rm Im}(d_{tG})$ are the same. All these aspects are checked 
to be satisfied 
within the statistical error of the event generation (100k events per point).
The quartic polynomial that fits these points is 
\begin{eqnarray}
\frac{\sigma}{\sigma_{SM}}&=& 
1 + 5.37\times 10^{-4}{\rm ~Im}(d_{tG})^2 + 2.41\times 10^{-8}{\rm ~Im}(d_{tG})^4 + 3.56\times 10^{-2}{\rm ~Re}(d_{tG})\nonumber \\
&+& 7.75\times 10^{-4}{\rm ~Re}(d_{tG})^2 + 5.87\times 10^{-6}{\rm ~Re}(d_{tG})^3 + 2.45\times 10^{-8}{\rm ~Re}(d_{tG})^4 \nonumber \\
\label{fit8}
\end{eqnarray}

\begin{figure}[thb]
\includegraphics[width=0.45\textwidth]{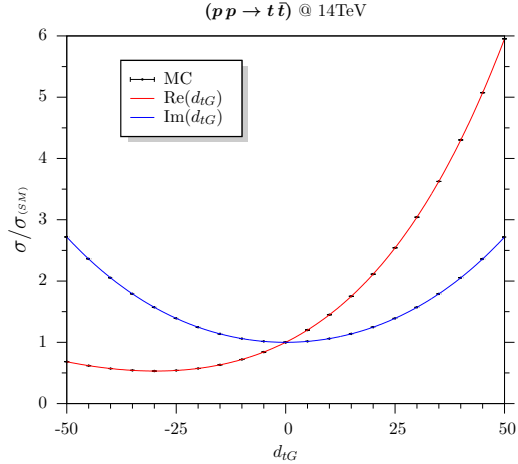}
\caption{Cross-section for the process $pp\to t\bar{t}$ at 14 TeV as calculated with {\tt MadGraph5} for different values of the anomalous coupling $d_{tG}$ and the corresponding fit.}
\label{dtgsig14}
\end{figure}

We next use {\tt MadGraph5} to generate event samples for the process $pp\to t\bar{t}$ at 14 TeV  for different values of the top CEDM and CMDM. The results of these numerical calculations, including the error estimated by {\tt MadGraph5}, are plotted in Figure~\ref{dtgsig14}. In the same figure we show a quartic fit to these results, as was described above for the 8~TeV case.
The resulting fit is given by
\begin{eqnarray}
\frac{\sigma}{\sigma_{SM}}&=& 
1 + 6.06\times 10^{-4}{\rm ~Im}(d_{tG})^2 + 3.28\times 10^{-8}{\rm ~Im}(d_{tG})^4 + 3.58\times 10^{-2}{\rm ~Re}(d_{tG})\nonumber \\
&+& 8.46\times 10^{-4}{\rm ~Re}(d_{tG})^2 + 6.76\times 10^{-6}{\rm ~Re}(d_{tG})^3 + 3.26\times 10^{-8}{\rm ~Re}(d_{tG})^4 \nonumber \\
\label{fit14}
\end{eqnarray}

Next we generate larger samples of $10^6$ events in the di-muon channel, $pp\to t\bar{t}\to b\bar{b}\mu^+\mu^-\nu\bar\nu$, for 4 values of ${\rm Im}(d_{tG})$ with the cuts of Eq.~\ref{cuts1} and the missing $\slashed{E}_T>30$~GeV requirement. These larger samples are necessary to estimate asymmetries at the $10^{-3}$ level. We use the four points to fit a linear form for the asymmetry since it can only be generated by the interference between CP conserving and CP violating amplitudes. The results of our event simulations and the corresponding fit are shown in Figure~\ref{dtgasym}. In principle the results could have terms cubic in the CEDM but our fits indicate their effect is negligible in this range of ${\rm Im}(d_{tG})$. One would expect the asymmetry to be generated by terms of the form ${\rm Im}(d_{tG})~{\rm Re}(d_{tG})$ which we have not simulated in our MC studies because they are expected at a much lower level. We have also checked that several other asymmetries that should vanish (for example those 
linear in the beam momentum) are consistent with zero within the statistical error of our event simulation. Finally, we have also verified numerically that 
there are no asymmetries induced by ${\rm Re}(d_{tG})$, as expected.

In Ref.~\cite{Gupta:2009wu} it was found that (Eq.~13 of that reference but with the notation of this paper) $A_1 \approx   - 4\times 10^{-3}  {\rm ~Im}(d_{tG})$ which is in reasonable agreement with our present result but not identical to it. There are two main differences between the two calculations: the different set of parton distribution functions; and the {\tt MadGraph5} implementation. Here we  implement the Lagrangian into {\tt MadGraph5} with the aid of {\tt FeynRules}, whereas in  Ref.~\cite{Gupta:2009wu} we directly modified the {\tt MadGraph4} code to insert the analytical result for the interference between the SM and the CEDM from Ref.~\cite{Antipin:2008zx}. The latter explicitly removes all terms that are not linear in the CEDM.

\section{$pp\to t\bar{t} h$ events}

\begin{figure}[thb]
\includegraphics[width=0.45\textwidth]{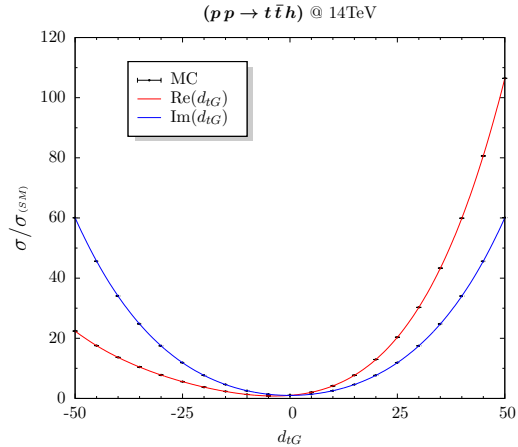}
\caption{Cross-section for the process $pp\to t\bar{t}h$ as calculated with {\tt MadGraph5} for different values of the anomalous coupling $d_{tG}$ and the corresponding fit.}
\label{dtgsigh}
\end{figure}

We first consider the production of a Higgs boson associated with a top pair, $pp\to t\bar{t}h$, and generate MC samples of $10^5$ events at $\sqrt{S}=14$~TeV for several  values of the top CEDM and the top CMDM. We show the results for the cross-sections in Figure~\ref{dtgsigh}.
We show in the same figure the result of a quartic fit to these points for extrapolation purposes. The number of parameters in the fit is reduced by noting that the cross-section only has quadratic and quartic terms for the CEDM resulting in
\begin{eqnarray}
\frac{\sigma}{\sigma_{SM}} &\approx&
1 + 1.53\times 10^{-2}{\rm ~Im}(d_{tG})^2 + 3.32\times 10^{-6}{\rm ~Im}(d_{tG})^4 + 1.14\times 10^{-1}{\rm ~Re}(d_{tG})\nonumber \\
&+& 1.70\times 10^{-2}{\rm ~Re}(d_{tG})^2 + 2.90\times 10^{-4}{\rm ~Re}(d_{tG})^3 + 3.32\times 10^{-6}{\rm ~Re}(d_{tG})^4 \nonumber \\
\label{sigtth}
\end{eqnarray}
The cross-sections obtained with this approximate formula are in agreement with the results in Ref.~\cite{Degrande:2012gr} for the linear and quadratic terms.

We next consider several asymmetries by generating samples of $10^6$ events (to reach a sensitivity of $10^{-3}$) for di-muon decays of the top pair and using the same cuts as in th previous section. The largest asymmetries we find are  
shown in Figure~\ref{f:tth-asym}. In addition we tried the operator
\begin{eqnarray}
{\cal O}_1 &=&  \vec{p}_h\cdot (\vec{p}_{\mu^+}\times\vec{p}_{\mu^-}) 
\end{eqnarray}
but it results in an symmetry too small to measure with $10^6$ events. We also tested correlations that are known to be zero, such as $ \vec{p}_{beam}\cdot (\vec{p}_{\mu^+}\times\vec{p}_{\mu^-}) $, to check that we indeed get zero within our statistical uncertainty. In the same manner we found that the CMDM does not induce any of the T-odd correlations, as expected.

\section{$pp \to b\bar{b}h$}

\begin{figure}[thb]
\includegraphics[width=0.45\textwidth]{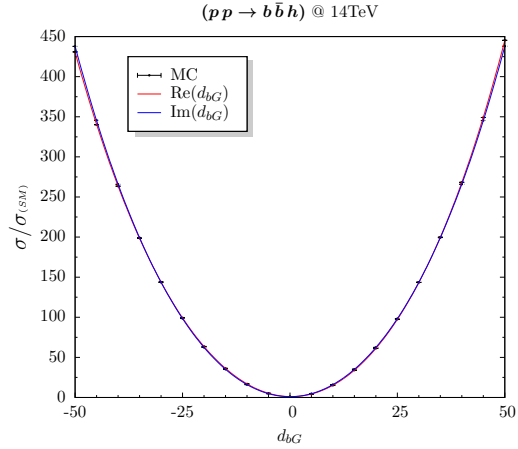}
\caption{Cross-section for the process $pp\to b\bar{b}h$ in the di-muon channel as calculated with {\tt MadGraph5} for different values of the anomalous coupling $d_{bG}$ and the corresponding fit.}
\label{dbgsigh}
\end{figure}

We repeat the exercise of the previous section, this time for the process $pp\to b\bar{b}h$. To this end we implement the $b$-quark CEDM and CMDM into {\tt MadGraph5} and compute the resulting cross-section for several values of these couplings. The results of these runs along with a quartic fit to them is shown in Figure~\ref{dbgsigh}.
The quartic fit to these points enforcing the condition of no odd powers of the CEDM is given by,
\begin{eqnarray}
\frac{\sigma}{\sigma_{SM}} &\approx&
1 + 1.49\times 10^{-1}{\rm ~Im}(d_{bG})^2  + 1.02\times 10^{-5}{\rm ~Im}(d_{bG})^4 - 9.76\times 10^{-2}{\rm ~Re}(d_{bG}) \nonumber \\
&+& 1.49\times 10^{-1}{\rm ~Re}(d_{bG})^2 + 9.73\times 10^{-5}{\rm ~Re}(d_{bG} )^3 + 1.02\times 10^{-5}{\rm ~Re}(d_{bG})^4 \nonumber \\
\label{sigbbh}
\end{eqnarray}

\section{$pp \to hX$}

To constrain the anomalous couplings $a_q^g$ and $d_q^g$ of the light quarks we consider their contribution to inclusive Higgs production. The dominant contribution is from the up-quark through the parton level diagrams $ug\to uh$ and $u\bar{u}\to gh$. Since we are not requiring a jet in the final state we have removed the $(p_T)_{min}$ cut associated with the final state quark (gluon) from the  {\tt MadGraph5} default cuts. We compute the resulting cross-section for several values of the couplings along with a quadratic fit(because in this case the new coupling appears only once in the amplitudes)  for the LHC at  $\sqrt{S}=8$~TeV.

The quadratic fit we obtain for these points is consistent with having no linear term for either the CEDM (as expected from CP) or the CMDM. The latter is consistent with the SM contribution being very small in the diagrams with up-quarks as it is proportional to the quark mass. The leading SM contribution to the tree level processes we evaluate arises from the parton level process $c\bar{c}\to hg$ and therefore only interferes with the charm quark CMDM. The result of the fit is given in Eq.~\ref{pphxfit}. 

\begin{figure}[thb]
\includegraphics[width=0.45\textwidth]{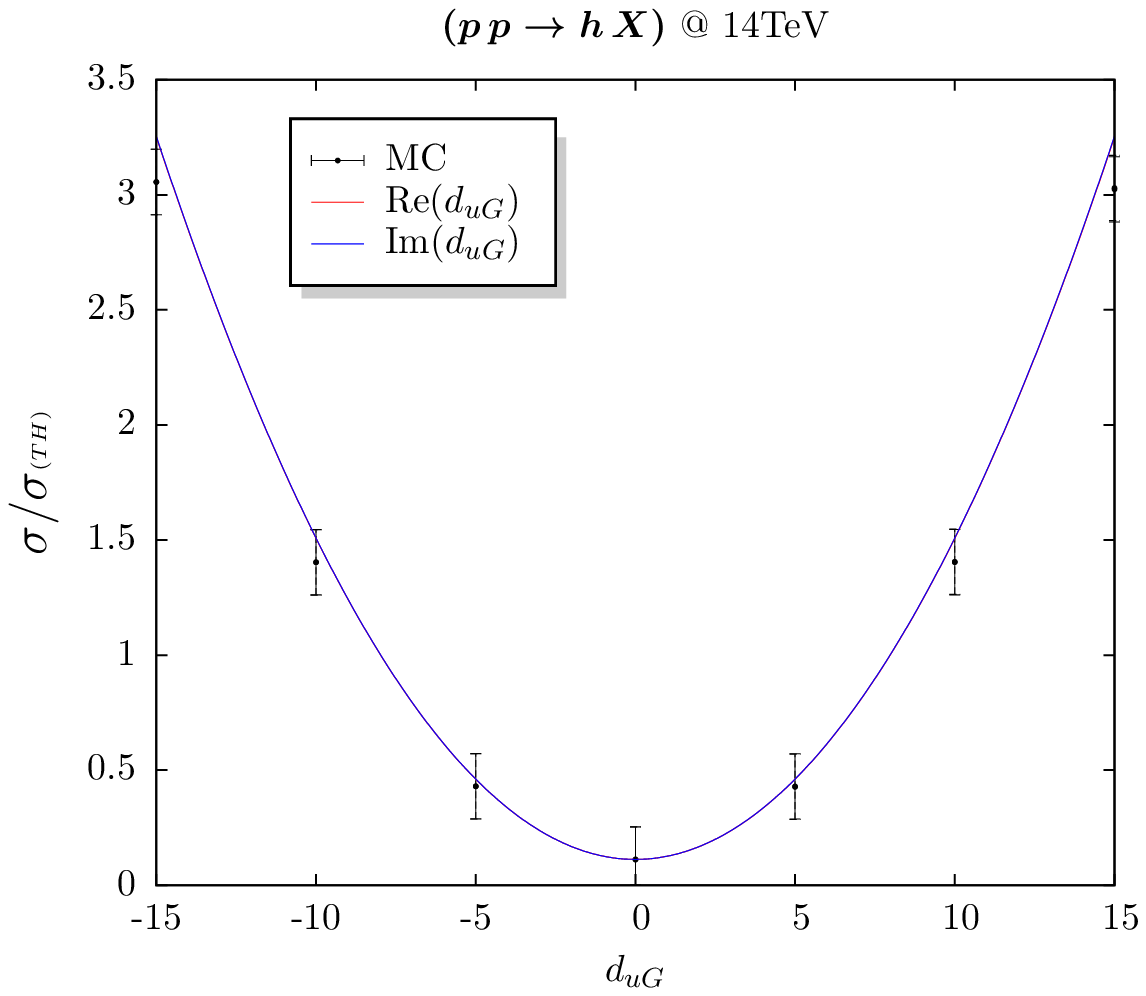} \hspace{1cm}
\includegraphics[width=0.45\textwidth]{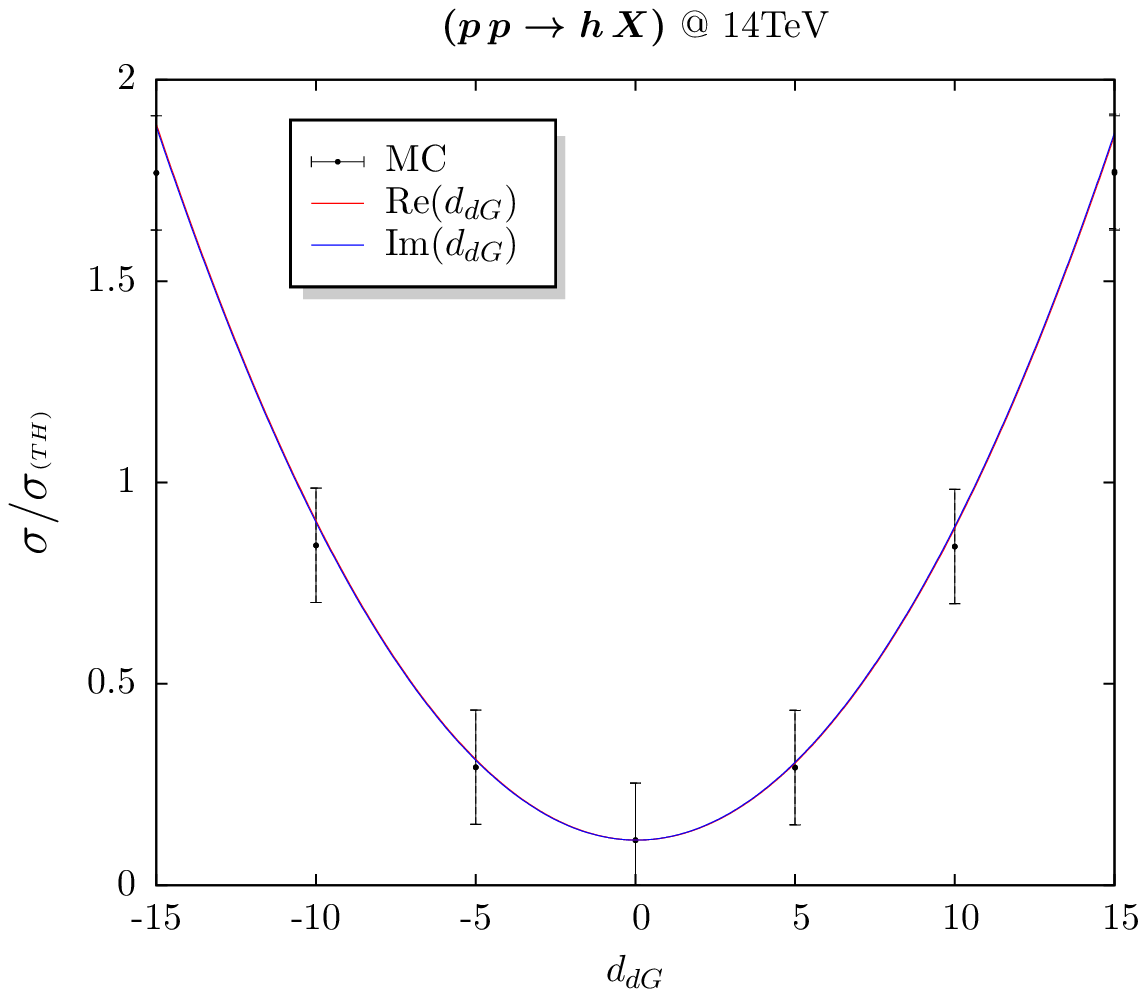} \\ \vspace{0.5cm}
\includegraphics[width=0.45\textwidth]{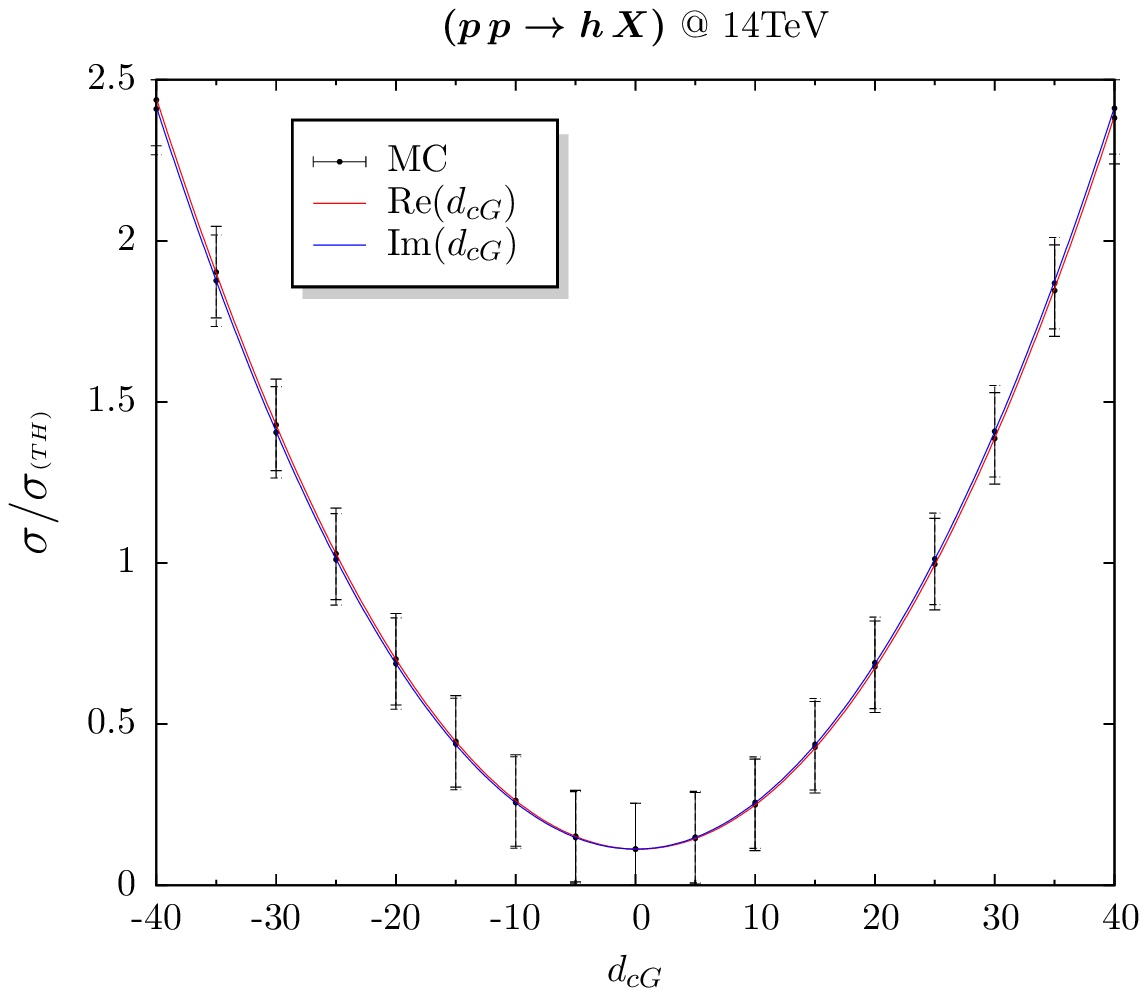} \hspace{1cm}
\includegraphics[width=0.45\textwidth]{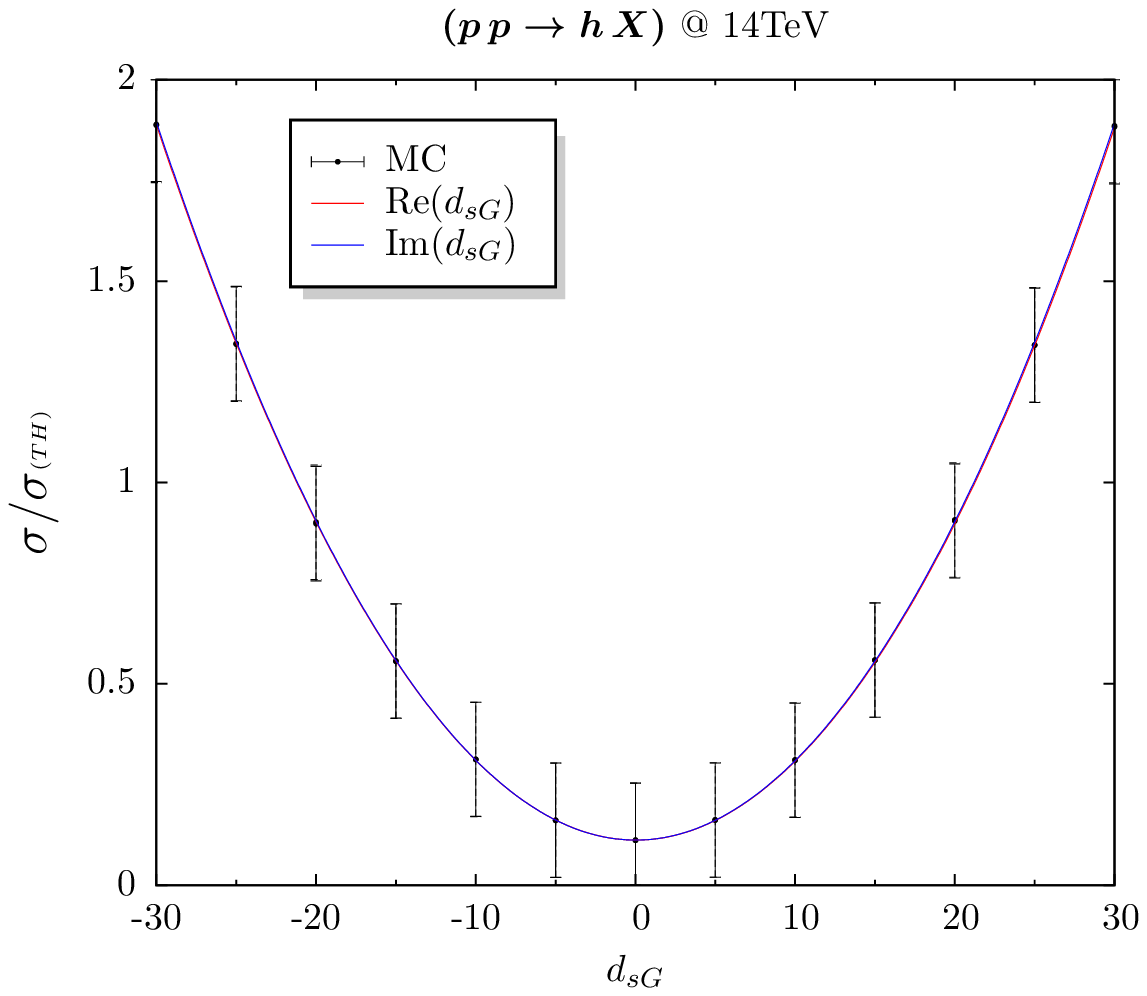} \\
\caption{Cross-section for the process $pp\to HX$ as calculated with {\tt MadGraph5} for different values of the anomalous coupling $d_{qG}$  and the corresponding fit  at  $\sqrt{S}=14$~TeV.}
\label{dugsigh14}
\end{figure}

The corresponding results for  the LHC at  $\sqrt{S}=14$~TeV are shown in Figure~\ref{dugsigh14}.
The resulting fit is given in Eq.~\ref{fithj14}.

\end{document}